\documentclass[reprint,twocolumn,superscriptaddress,showpacs,nofootinbib,notitlepage,floatfix]{revtex4-1}
\usepackage[utf8]{inputenc}
\usepackage{graphicx}
\usepackage{latexsym,amsmath,amssymb,lmodern,float,url}
\usepackage{natbib}
\usepackage{color}
\usepackage{microtype}
\usepackage{slashed}
\usepackage{multirow}
\usepackage{qcircuit}
\usepackage{tikz}
\usepackage{xcolor}
\usepackage{float}

\usepackage[colorlinks=true,backref=false, linktocpage=true,
citecolor=blue,urlcolor=blue,linkcolor=blue,pdfpagemode=UseOutlines]{hyperref}

\hypersetup{%
  bookmarksnumbered=true,
  pdftitle = {},
  pdfsubject = {},
  pdfauthor = {},
  pdfkeywords = {}
}

\newcommand{\eye}{I}

\begin{document}
\title{Quantum algorithms for disordered physics}

\author{Andrei Alexandru}
\email{aalexan@gwu.edu}
\affiliation{Department of Physics, The George Washington University, Washington, D.C. 20052, USA}
\affiliation{Department of Physics, University of Maryland, College Park, MD 20742, USA}
\author{Paulo F. Bedaque}
\email{bedaque@umd.edu}
\affiliation{Department of Physics, University of Maryland, College Park, MD 20742, USA}
\author{Scott Lawrence}
\email{srl@umd.edu}
\affiliation{Department of Physics, University of Maryland, College Park, MD 20742, USA}
\date{\today}

\begin{abstract}
We show how a quantum computer may efficiently simulate a disordered Hamiltonian, by incorporating a pseudo-random-number generator directly into the time evolution circuit. This technique is applied to quantum simulation of few-body disordered systems in the large volume limit, in particular Anderson localization. The method requires a number of (error corrected) qubits proportional to the logarithm of the volume of the system, and each time evolution step requires a number of gates polylogarithmic in the volume. We simulate the method to observe the metal-insulator transition on a three-dimensional lattice. Additionally, we demonstrate the algorithm on a one-dimensional lattice, using physical quantum processors.
\end{abstract}

\maketitle


Random potentials can cause quantum particles to localize, that is, to remain in a finite region of space instead of diffusing~\cite{Anderson:1958vr,doi:10.1142/7663,RevModPhys.80.1355}. In three dimensions and at fixed energy (and for specific lower-dimensional systems), this effect only occurs after some critical disorder is reached, resulting in a second-order phase transition --- the Anderson localization transition~\cite{PhysRevLett.42.673}. Physically, this is a transition between a conductor and an insulator. Numerical access to this transition is hampered by the large Hilbert spaces needed in the limit of large volumes. Additionally, it is unclear to what extent the transition is affected by the presence of interactions~\cite{PhysRevB.21.2366,PhysRevB.37.325}.
Localization is a purely quantum mechanical phenomenon  caused by interference among different paths. It is perhaps not surprising then that a quantum computer is particularly well suited to capture its physics. The purpose of this paper is to flesh out a method to study localization in quantum computers with a number of qubits scaling as the logarithm of the size of the lattice, opening up the prospect of useful physical simulation with a small number of (corrected) qubits.

The production of small-scale quantum computers has created great interest in their possible physical applications. Quantum computers are able, not just in principle but in practice~\cite{Arute:2019}, to perform computations infeasible on a classical computer. 
The simulation of fundamentally quantum systems is a particularly natural target for quantum computers~\cite{Lloyd:1996}. Fault-tolerant qubits, once practical, will allow the use of a wide variety of algorithms designed without regard for the particular implementation of qubits and gates used by the hardware. However, many algorithms proposed for simulating physical systems require large numbers of qubits before becoming physically relevant~\cite{Klco:2018zqz,Zohar:2018cwb,Zohar:2014qma,Roggero:2018hrn,haah2018quantum,Jordan:2014tma,Jordan:2011ne,Lamm:2019bik,Zache:2018jbt,Martinez:2016yna,Yeter-Aydeniz:2018mix,Alexandru:2019ozf}, which will not be available for the foreseeable future. In the systems that give rise to Anderson transitions, the dimension of the Hilbert space increases only linearly with the volume, suggesting that a logarithmic number of qubits could be used, and physically relevant systems could be attacked with near-term resources.

In this paper we detail algorithms for the quantum simulation of the Anderson transition, and demonstrate their operation on a simulated digital quantum computer. These algorithms generically require only a logarithmic number of qubits in the volume of the physical system, corresponding to an exponential speedup over equivalent classical algorithms. The number of qubits needed to make the output of these algorithms physically interesting is only $\sim 50$, recommending this method as a target for near-term fault-tolerant quantum computers.

The Hamiltonians we will consider simulating contain $O(V)$ terms of random amplitudes (the source of the disorder). Selecting these values classically would require passing $O(V)$ values to the classical computer, precluding the possibility of $O(\log V)$ efficiency. The core of our method is to pass the quantum computer not a list of values, but instead a small circuit (actually a PRNG, a pseudo-random number generator) which maps an integer $i$ to the pseudo-random variable $u_i$, interpreted as a coefficient in the Hamiltonian. This circuit defining the Hamiltonian, is incorporated directly into the time-evolution operation.

We begin by considering the Anderson tight-binding model of a single particle hopping between $V$ lattice sites:
\begin{equation}\label{eq:site-disorder}
H_{\mathrm{site}} = - \sum_{<ij>} \left(c^\dagger_i c_j + \mathrm{h.c.}\right)
+
W \sum_i u_i c^\dagger_i c_i
\text.
\end{equation}
Here the first sum is taken over all adjacent sites. The random variables $u_i$ are uniformly distributed in $[0,1)$ and are time independent. The only tunable parameter is $W$, the disorder parameter. In the limit $W = 0$, this is a free theory, in which all eigenstates are delocalized.

Macroscopic properties of this model --- those apparent in the limit of large volumes --- do not depend on the particular instantiation of $u_i$. The most interesting such property is localization: in one and two dimensions, any non-vanishing disorder $W$ causes all eigenstates of the Hamiltonian to be localized, and the medium becomes an insulator. In three dimensions, the medium remains a conductor up to a critical disorder $W_c \approx 16.5$~\cite{brandes2003anderson}, above which all states are again localized. This phase transition in three dimensions is an easy target for quantum simulation.

\begin{figure}
\centerline{
\Qcircuit @C=1.2em @R=0.6em {
& \qw & \targ{2} & \qw & \qw & \qw & \qw & \qw & \targ & \qw\\
& \qw & \ctrl{-1} & \targ & \qw & \qw & \qw & \targ & \ctrl{-1} & \qw\\
& \gate{R_X(\Delta t)} & \ctrl{-2} & \ctrl{-1} & \targ & \gate{R_X(\Delta t)} & \targ & \ctrl{-1} & \ctrl{-2} & \qw\\
}
}
\vspace{1.5em}
\centerline{
\Qcircuit @C=1.2em @R=0.6em {
& \multigate{2}{P} & \gate{R_Z(\frac{W \Delta t}{2})} & \multigate{2}{P^\dagger} & \qw \\
& \ghost{P} & \gate{R_Z(\frac{W \Delta t}{4})} & \ghost{P^\dagger} & \qw \\
& \ghost{P} & \gate{R_Z(\frac{W \Delta t}{8})} & \ghost{P^\dagger} & \qw \\
}
}
\caption{Time-evolution circuits for the Anderson tight-binding model with site disorder, on eight sites. On top is shown approximate evolution under the kinetic Hamiltonian $H_K$; on the bottom is the evolution under the disordered potential $H_V$. The construction of the random permutation $P$ is discussed in the text.\label{fig:circuits}}
\end{figure}
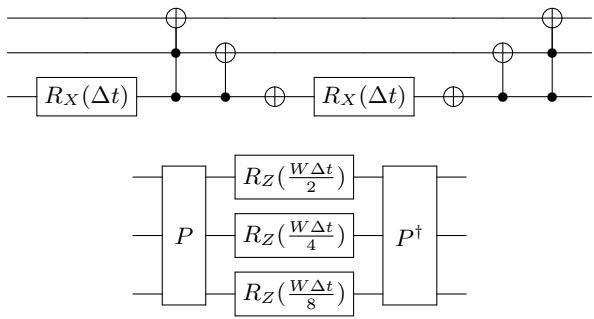

We first discuss the proposal in the context of one-dimensional model and then indicate how to extend it to higher dimensions.
To simulate the Anderson tight-binding model in a single dimension, interpret each element of the computation basis as a binary integer specifying the position of the site ($0\ldots (L-1)$) occupied by the electron. Thus, $\left|01101\right>$ is the state of the system when the electron is at site $13$ --- we will abbreviate this state as $\left|13\right>$. The time evolution
 is Trotterized, and the effects of the kinetic $H_K = - \sum_{<ij>} \left(c^\dagger_i c_j + \mathrm{h.c.}\right)$ and potential $H_V = W \sum_i u_i c^\dagger_i c_i$ terms are implemented sequentially:
 \begin{equation}
 e^{-i Ht} \approx  \left(  e^{-i H_K \Delta t}  e^{-i H_V \Delta t}  \right)^{t/\Delta t}.
 \end{equation}

In this formulation of the Anderson model, the kinetic Hamiltonian $H_K$ could be diagonalized by a quantum Fourier transform; however, this is no longer the case when the disorder lies on the links, as in a model discussed below. Because of this we instead simulate the evolution of $H_K=H_{K,e}+H_{K,o} $ by splitting it into even ($H_{K,e}$) and odd ($H_{K,o}$) links. Since $H_{K,e}$ and $H_{K,o}$ commute there is no error arising from this Trotterization.
The Hamiltonian of the even links couples sites that differ only in the last qubit
\begin{equation}
H_{K,e} = \left|0\right>\left<1\right| + \left|2\right>\left<3\right| + \cdots + \mathrm{h.c.} = \eye^{\otimes (V-1)} \otimes \sigma_x
\text,
\end{equation}
and $e^{-i H_{K,e} \Delta t}$ is therefore simulated by a rotation about the $X$ axis of the least-significant qubit.
The evolution of the odd links appears more complicated, but is simplified by first translating the entire lattice by $1$.  By this translation, odd links are transformed into even links. After the shift, time-evolution under $H_{K,e}$ is performed again, followed by a shift  back into place. The resulting circuits for evolving $e^{-i H_K \Delta t}$ are shown in Figure~\ref{fig:circuits}.

Having evolved with $H_K$, we come to the crux of our proposal: evolving under the disordered potential
\begin{equation}
H_V = W \sum_i u_i \left|i\right>\left<i\right|\text.
\end{equation}
The evolution under $H_V$ requires the phase of the state to be changed by the same random number
every time the particle finds itself at a particular lattice site (which corresponds to a state in the computational basis).  That means that the same random number has to be generated multiple times.
 Determining the random values $u_i$ classically and constructing a circuit using them would necessitate $O(V)$ gates in the time-evolution, but with a different approach we can obtain a cost merely polylogarithmic in the volume. We first observe that with standard methods on a classical computer, the $u_i$ are typically not truly random variables, but instead are defined to be the output of a pseudo-random number generator (PRNG) --- a circuit sufficiently complicated so that the $u_i$ look random to any practical statistical test.  Therefore, we will define $H_V$ to be specified by the output of a particular \emph{seekable} PRNG, and use that circuit directly in the time evolution.

A seekable PRNG is a PRNG from which the $i$th element $f(i)$ can be obtained in constant time, independent of the value of $i$. This contrasts with the PRNGs used in most computer simulations, in which elements must be calculated sequentially, so that calculating $f(i)$ requires $O(i)$ steps. Given a short (logarithmic in the number of bits $Q$ used to represent $i$) classical circuit computing a PRNG $f$, one can readily produce a quantum circuit implementing a unitary $U_f$ defined by $U_f \left|i\right>\left|0\right> = \left|i\right>\left|f(i)\right>$. Taking the output of the PRNG to define the site disorder by $u_i = 2^{-Q} f(i)$, so that
\begin{equation}
H_V = W \sum_i \frac{1}{2^Q} f(i) \left|i\right>\left<i\right|
\text,
\end{equation}
we see that evolution under $H_V$ can be obtained by first applying $U_f$ to compute the PRNG, performing a diagonal phase rotation by the amount specified by the value in the anciliary register, and then applying $U_f^\dagger$.

The setup above requires the construction of a suitable $f(i)$. Seekable PRNGs are often defined via cryptographic hash functions like {\tt SHA256}~\cite{sha256ref}. In this construction, a sequence of $K$ PRNGs indexed by seeds $k$ are constructed via $f_k(i) = {\tt SHA256}(i*K+k)$. This construction is known to perform very well~\cite{salmon:2011}, but cryptographic hashes typically operate on fixed-size registers of large numbers of bits, and the well-studied ones are therefore unobtainable on near-term quantum computers. For this reason, we instead define a seekable PRNG with a random classical circuit --- that is, a random permutation matrix.

\begin{figure}
\centering
\includegraphics[width=0.99\columnwidth]{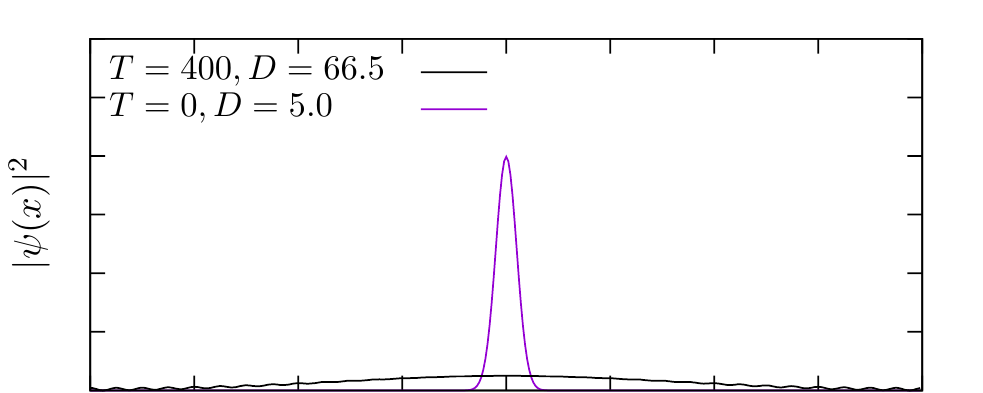}
\includegraphics[width=0.99\columnwidth]{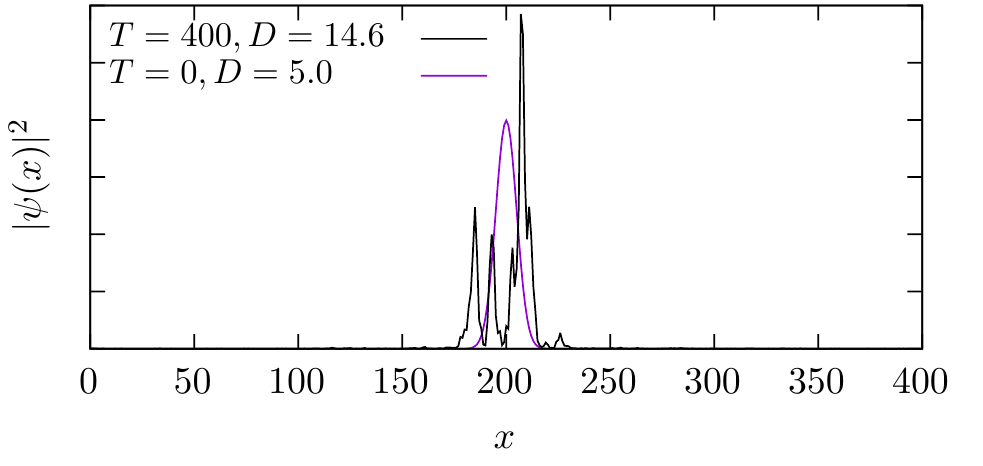}
\caption{Visualization of the diffusion (top) and localization (bottom) of a particle on a one-dimensional lattice with $400$ sites. In each plot the probability of the wavefunction is shown for two different times. Without disorder ($W=0$), the average distance given by Eq.~\ref{eq:coconut} diverges, and $D/L$ saturates. With disorder $W=1$, the wavefunction fails to spread out even at long times, and $D/L$ remains small. \label{fig:wavefunctions}}
\end{figure}

A random permutation matrix $P$ may be used to define a seekable PRNG via $P\left|i\right> = \left|f(i)\right>$. This introduces weak correlations between the different values of $f(i)$, as $f(i) \ne f(j)$ whenever $i\ne j$; these correlations are unmeasureable in the large volume limit, and we will ignore them.

It remains to construct a random permutation operator on a quantum computer. 
Our method will be to add randomly a number of \texttt{NOT} and \texttt{TOFFOLI} gates on an extended register with $Q+(Q-3)$ qubits. It is easy to see this generates a random permutation with a probability independent of the permutation. Indeed, 
with the aid of $Q-3$ extra anciliary qubits, any permutation matrix can be obtained by composing the \texttt{NOT} and \texttt{TOFFOLI} gates~\cite{Toffoli:1980}. Beginning with an empty circuit (corresponding to the identity), consider the process of appending one randomly-selected gate at a time (the probability of picking any of the \texttt{NOT} or \texttt{TOFFOLI} gates is irrelevant as long as it does not vanish for any of them). This is a Markov chain in the space of permutations which, by virtue of Toffoli's construction, is ergodic. As long as one of the gates included is the identity gate, it is also aperiodic, and therefore the chain converges to the unique stationary distribution where every permutation is equally probable. Therefore, a sequence of $G$ randomly selected \texttt{NOT} and \texttt{TOFFOLI} gates will converge to a random permutation matrix as $G \rightarrow\infty$. This is true regardless of the distribution of gates used, although the distribution does affect the rate of convergence.
As one additional simplification, for our demonstrations we dispense with the anciliaries which are required to sample a truly random permutation matrix. This modification makes no difference in the large-volume limit. To see this, relabel the highest-order half of the qubits, representing the least-significant digits, as ancillary. These qubits contributed very little to the energy level of a site, so this change disappears in the infinite-volume limit.

Although we have shown that with enough random gates this construction will yield random numbers, it is important that the number of gates required scales polynomially, rather than exponentially, with $Q$. To demonstrate that this is the case, we run the \texttt{dieharder}~\cite{dieharder} battery of statistical tests (based on the diehard tests~\cite{diehard}) on random numbers generated by random classical circuits on $30$ bits. As \texttt{dieharder} takes in a sequence of $8$-bit numbers, we truncate each random number to the highest-order $8$ bits. Trying different lengths of gates, we find that $600$ gates is sufficient to consistently pass all statistical tests in the battery. Similarly, for random classical circuits on $50$ bits, we find $1200$ gates to be sufficient, and for $100$ bits $2600$ gates. This is suggestive of low-exponent polynomial scaling.

The resulting circuit for evolving a three-qubit (eight-site) model under $H_V$ is shown in Figure~\ref{fig:circuits}.

The generalization of our algorithm to higher dimensions is straightforward.
The random potential in the Hamiltonian is local and makes no reference to the spatial structure of the lattice, and so that portion of the time evolution is unaffected by the number of dimensions. The kinetic Hamiltonian factorizes, and we have in three dimensions
\begin{equation}
H_K^{(3)}
=
H_K \otimes \eye \otimes \eye
+
\eye \otimes H_K \otimes \eye
+
\eye \otimes \eye \otimes H_K
\text,
\end{equation}
so that kinetic evolution may be performed independently in each dimension separately.

\begin{figure}
\centering
\includegraphics[width=0.99\columnwidth]{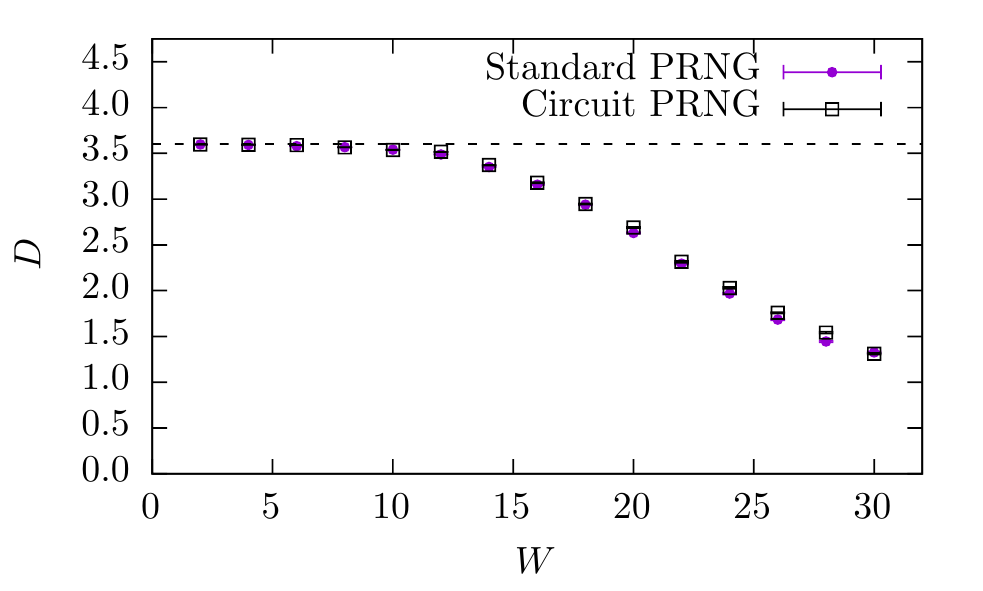}
\caption{The average distance, measured by Eq.~\ref{eq:coconut}, as a function of the disorder parameter $W$, for a $16^3$ lattice. The blue triangles show results obtained with a conventional PRNG, and the red squares give results obtained with the circuit-based PRNG described here. The localization distance $D$ is measured at asymptotically large times and the stochastic error bars are generated using $300$ samples. The dashed line represents the delocalized limit for $D$. \label{fig:site-transition}}
\end{figure}

This completes the construction, on a digital quantum computer, of the time evolution operator $e^{-i H t}$. Many experiments can be performed once this time-evolution is available. The most natural is to begin with an electron at a particular site ($\left|\Psi(t=0)\right> = \left|0\right>$) and measure its propagation in time. The two possible qualitative behavior can be seen in Figure~\ref{fig:wavefunctions}. On a quantum computer, by beginning in that state, time-evolving, and measuring the position, we can then measure the observable
\begin{equation}\label{eq:coconut}
D = \frac{L}{\pi\sqrt{2}}\sqrt{1 - \left<\cos \frac{2\pi \hat x}{L}\right>}
\text,
\end{equation}
which is equivalent in the infinite-volume limit to the average distance from the origin measured along the $x$-direction.
In the insulating phase, in the long time limit, the electron will remain near site $0$ and $D$ will asymptote to a constant; in the conducting phase, it will diffuse and eventually cover the lattice, setting the value of $D$ to its maximum $D=L/(\pi\sqrt{2}) $. Thus, preparing the state $\left|0\right>$ (or any other basis state), evolving for a long time, and finally measuring the position of the electron, informs us which phase we are in. In practice, we plot $D/L$, which is $0$ in the infinite volume limit of the insulating phase and a finite constant in the conducting phase. Such a plot, computed classically, is shown in Figure~\ref{fig:site-transition}; note that the results with random circuits (as discussed above) agree with those obtained using a conventional PRNG.

\begin{figure}[t]
\includegraphics[width=0.99\columnwidth]{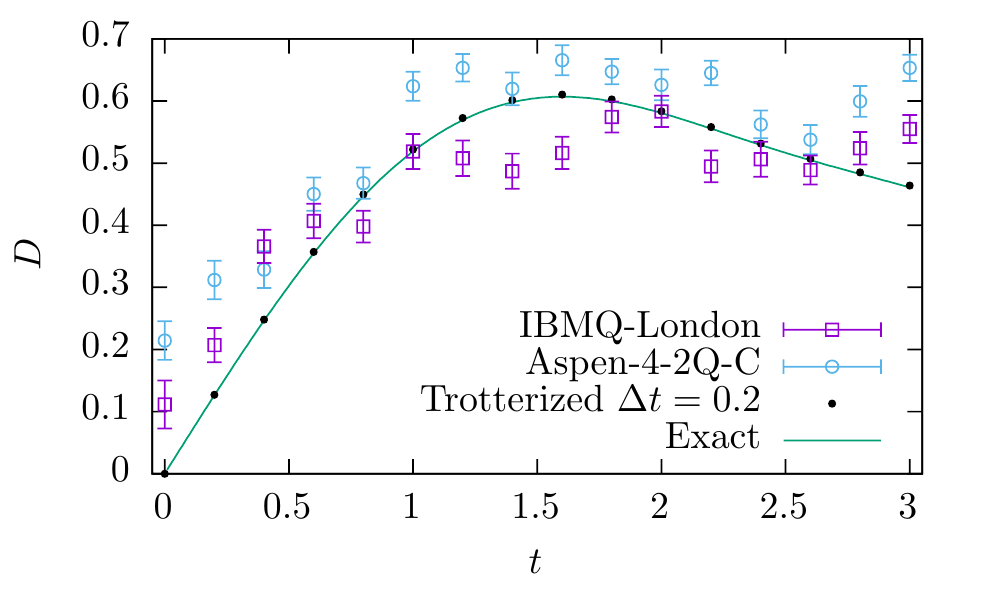}
  \caption{Average distance, as measured by Eq.~\ref{eq:coconut}, as a function of evolution time for a $4$-site lattice with disorder parameter $W=5$ and Trotterization size $\Delta t = 0.2$. The ``random'' potential is fixed to be $V(0,1,2,3) = 0,1,3,2$, which may be computed with a single \texttt{CX} gate. The data points are obtained on physical quantum processors provided by IBM and Rigetti and are compared with the exact result. We perform 300 quantum measurements for each point to estimate $D(t)$ and the stochastic errors shown are computed using bootstrap.\label{fig:qsim}}
\end{figure}

This method cannot yet be implemented at scale on a quantum computer: achieving quantum advantage would require about $50$ fault-tolerant qubits, which are not yet available. The state of what is currently achievable is depicted in Figure~\ref{fig:qsim}. With publicly available quantum computing facilities provided by Rigetti, we are able to simulate around $10$ time-steps with $4$ sites before noise and decoherence entirely take over.


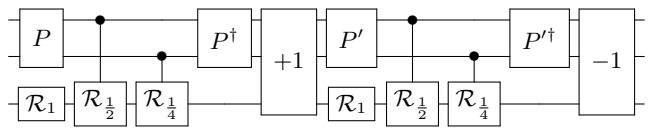
\begin{figure}[b]
\vskip 1cm
\centering
\resizebox{0.99\columnwidth}{!}{
\Qcircuit @C=0.4em @R=0.6em {
& \multigate{1}{P} & \ctrl{2} & \qw & \multigate{1}{P^\dagger} & \multigate{2}{+1} & \multigate{1}{P'} & \ctrl{2} & \qw &  \multigate{1}{P'^\dagger} & \multigate{2}{-1} & \qw\\
& \ghost{P} & \qw & \ctrl{1} & \ghost{P^\dagger} & \ghost{+1} & \ghost{P'} & \qw & \ctrl{1} & \ghost{P'^\dagger} & \ghost{-1} & \qw\\
& \gate{\mathcal R_{1}} & \gate{\mathcal R_{\frac 1 2}} & \gate{\mathcal R_{\frac 1 4}} & \qw & \ghost{+1} & \gate{\mathcal{R}_1} & \gate{\mathcal R_{\frac 1 2}} & \gate{\mathcal R_{\frac 1 4}} & \qw & \ghost{-1} & \qw\\
}
}
\caption{Circuit for evolving the link-disorder Hamiltonian of Eq.~\ref{eq:link-disorder} on a three-qubit, eight-site system. For brevity, we use $\mathcal R_\theta = R_X(\theta \Delta t)$, and the increment and decrement circuits are not shown explicitly. Note that $P$ and $P'$ are independently sampled permutation matrices.\label{fig:link-circuits}}
\end{figure}

The site disorder of Eq.~\ref{eq:site-disorder} is not the only way to introduce disorder onto the lattice. Another option is to introduce the disorder on the hopping terms:
\begin{equation}
\label{eq:link-disorder}
H_{\mathrm{link}} = \sum_{\left<ij\right>} \left(W u_{ij} - 1\right)\left[
c^\dagger_i c_j + \mathrm{h.c.}
\right]
\end{equation}
In this model the link between $i$ and $j$ has an associated random variable $u_{ij}$, and there is no potential on the sites. To evolve this Hamiltonian we may use a similar strategy as before. In one dimension, two random permutation matrices on $Q-1$ qubits are constructed. They determine the disorder on even and odd links, respectively. The circuit for performing evolution under this Hamiltonian is shown in Figure~\ref{fig:link-circuits}.

We have discussed site-disorder and link-disorder on cubic lattices of arbitrary dimension. In general, a Hamiltonian which is sparse and efficiently row-computable (a short circuit can be presented to compute the non-zero entries in any row of the Hamiltonian) can be simulated in time logarithmic in the dimension of the physical Hilbert space~\cite{Aharonov:2003}. The trick used in this paper is quite general: when considering a model with many random values in the Hamiltonian, those values may be defined to be the output of a PRNG without changing the macroscopic properties of the model. This removes the need to specify $O(V)$ values explicitly in a circuit, and allows efficient simulation of the Hamiltonian.

The algorithms presented here allow the physics of metal-insulator transitions to be accessed with a number of qubits scaling with the logarithm of the volume of the system. These do require fault-tolerant qubits, which have not yet been demonstrated in practice. However, about $50$ such qubits would already allow quantum computers to push past the systems that can be treated with a classical computer.

Besides going to larger volumes than a classical computer could hope to achieve, the gentle scaling of resources with Hilbert space opens up the possibility of studying few-body interactions in random potentials as well as Anderson transitions in higher dimensions, with an eye toward determining the upper critical dimension~\cite{Tarquini:2017}.

\begin{acknowledgments}
We are grateful to Tom Cohen and Brian Swingle for their suggestions on the course of this work, and to Jay Sau for comments on an earlier version of this manuscript.
A.A. is supported in part by the U.S. Department of Energy grant DE-FG02-95ER-40907. P.B.\ and S.L.\ are supported by the U.S. Department of Energy under Contract No.~DE-FG02-93ER-40762.
\end{acknowledgments}


\bibliographystyle{apsrev4-1}
\bibliography{anderson}
\end{document}